\documentclass[a4paper,11pt]{article}
\usepackage{jcappub} 

\usepackage[T1]{fontenc} 

\title{Hubble multi-scalar inflation}

\author{Habib Abedi}
\author{and Amir M. Abbassi}
\affiliation{Department of Physics, University of Tehran,\\North Kargar Ave, Tehran, Iran.}
\emailAdd{h.abedi@ut.ac.ir} 
\emailAdd{amabasi@khayam.ut.ac.ir} 
\abstract{
 Multiple field models of inflation exhibit new features than single field models. In this  work, we study the hierarchy of parameters based on Hubble expansion rate in curved field space and derive the system of flow equations that describe their evolutions.
Then we focus on obtaining derivatives of number of $e$-folds with respect to scalar fields during inflation and at hypersurface of the end of inflation.}
\arxivnumber{?}
\begin{document}
\maketitle
\flushbottom
 
\section{Introduction}

Inflationary paradigm is the  prevalent  description of the early universe. 
This provides an almost scale invariant and nearly Gaussian spectrum of adiabatic cosmological perturbations.
 We have evidence in favor of single field models \cite{Planck}, but there are good reasons to search beyond this picture and consider multiple active scalar fields during inflation.
 
Multiple field models of inflation have been extensively studied  \cite{Multifield1, Multifield2,Multifield3, Multifield4, Multifield5, Emery13} and show a number of interesting features. Curvature perturbations in single field models are constant outside of the horizon in all orders of perturbation. However it can change in multi-field ones due to entropy perturbations, unless background trajectory follows a geodesic of the field space
 \cite{Nibbelink,Langlois,Peterson}. 
Multi-field models can also make considerable non-Gaussianity \cite{NG1, NG2,NG3}.
In addition,
 there are some effects at the end of inflation.
 In multiple field models, inflation does not end at the same values of fields at different places.
Thus the hypersurface of the end of inflation is not uniform in multiple field models, so it can lead to different number of $e$-folds at different places. Different number of $e$-folds affect the curvature perturbations at the end of inflation \cite{End1, End2, End3, End4, End5, End6}.

In modified inflationary models Lagrangian can contain non-canonical kinetic terms. It can be an arbitrary function of canonical kinetic term which is known as k-inflation  \cite{Kinflation1, Kinflation2}.
Such modifications lead to  reduced sound speed for perturbations.
Other non-canonical forms of kinetic term can be in the form of $ -{\cal G}_{IJ}(\phi^K) \, \partial_\mu \phi^I \, \partial^\mu \phi^J/2 $, where ${\cal G}_{IJ}$   interpreted as the metric of  field space are only functions of scalar fields.
 String theory motivates us to use many scalar fields with non-canonical kinetic terms  \cite{String}.
Moreover, supersymmetric models of inflation  include nontrivial K\"{a}hler potentials which modify the field space metric.
 Scalar fields may also be non-minimally coupled to gravity \cite{Buchbinder}.
 Conformal transformations can convert  the Lagrangian with the non-minimal coupling  into the standard Einstein-Hilbert Lagrangian minimaly coupled, but with non-canonical kinetic terms \cite{Conformal, Conformal1}.

 A convenient approach to study the super-horizon scales is gradient expansion. The leading order approximation in the gradient expansion, i.e. $\delta N$-formalism or the separate universe approach, use only the background dynamics to calculate the curvature perturbation $\zeta$ on large scales \cite{Delta1, Delta2, Delta3, Sugiyama, Garriga}.
In $\delta N$-formalism, usually authors assume simplified cases, e.g. the assumption of sum-separable \cite{Wand24, Seery1, Easther2, Sasaki3, Bramante} or product-separable potentials \cite{Bellido, Choi}.

In Hamilton-Jacobi approach one treats the Hubble expansion rate as the fundamental quantity \cite{Salopek, Coone, Horner, Powell, Spalinski, Contaldi, Emery, L2}.
 The exact evolution of spacetime in presence of scalar fields can be described in terms of the rate of expansion and its derivatives w.r.t. scalar fields, which are functions  of scalar fields. 
We can define a set of useful parametrizations of inflation in this approach.
 Liddle \cite{L1}  for single field and Easther and Giblin \cite{Easther1} for multiple canonical scalar fields  have shown that in the case of  truncated parameters, flow equations  can be solved analytically.
Consequently, they used the hierarchy of parameters and derived the system of flow equations to find full inflationary evolution as a function of the number of $ e $-folds $N$. 
Moreover this formalism is convenient for study of inflation beyond slow-roll approximation \cite{Campo}. 
In ref.\cite{Tasinato1}, sum-separable Hubble parameter without considering slow roll approximation was studied. They obtained a class of inflationary solution and  three-point correlation function. 

The aim of this work is to study multi-field inflation in Hamilton-Jacobi formalism and to calculate the terms that  contribute in  $\delta N$-formalism during and at the end of inflation. In section \ref{Curved field-space} we review $\delta N$-formalism and the separate universe approximation.
Section \ref{Flow} is devoted to derive a system of first order differential equations of  inflation parameter in curved field space.
We obtain the same flow equations as flat field space with just replacing the covariant derivative in field-space instead of partial derivative. 
However the condition for truncating these parameters is more complicated than flat case. 
We use the $\delta N$ formalism in curved field space in section \ref{CP}.
Same calculations have been done with canonical kinetic terms that we extend it to non-canonical one. Our calculations can be done for other non-flat models. The derivatives of $N$ w.r.t. scalar fields play crucial role in this formalism and is central part of this section.
Section \ref{The adiabatic direction} is about perturbations in kinematic basis and finally last section is devoted to  scalar fields non-minimally coupled to gravity.

Throughout this work we use $ I,J,K, \cdots=1,2,\cdots,M $ for M scalar fields $ \phi^I $; $\mu, \nu, \cdots =1,\cdots, 4$ for spacetime indices and $i, j, \cdots=1,2,3$ for spatial coordinates. We also assume background spatially flat spacetime and we set $M_p^2=1$.

 \section{Gradient expansion}
\label{Curved field-space}

We write the action of $M$ scalar fields $\phi^I$ minimally coupled to standard Einstein-Hilbert's gravity as:
\begin{equation}
S=\frac{1}{2} \int d^4x \sqrt{-g} \left[ R- {\cal G}_{IJ}(\phi^K) \; \partial_\mu \phi^I \; \partial^\mu \phi^J-2V(\phi^K) \right],  \label{action1}
\end{equation}
where  $V(\phi^I)$ is general form of the potential. It is common to consider positive components of field space metric to avoid ghost instabilities.
The action of scalar fields with canonical  kinetic terms is recovered by $ {\cal G}_{IJ}=\delta_{IJ} $.
The field space metric is used to raise and lower field space indices.
The energy-momentum tensor of scalar fields gets the following form;
\begin{equation}
T_{\mu \nu}={\cal G}_{IJ} \, \partial_\mu \phi^I \, \partial_\nu \phi^J -  \frac{1}{2} g_{\mu \nu} \left[ {\cal G}_{IJ} \partial_\alpha \phi^I \, \partial^\alpha\phi^J + 2V(\phi^I) \right] .
\end{equation}
The Arnowitt-Deser-Misner (ADM) line element reads
\begin{equation}
ds^2=-{\cal L}^2 \, dt^2 +h_{ij} (dx^i+N^i \, dt) (dx^j+N^j \, dt).
\end{equation}
In this foliation the lapse function $ {\cal L} $ and shift vector $ N^i $ are Lagrange multipliers and $ h_{ij} $ is  spatial three-metric which contains dynamical degrees of freedom.
Plugging the ADM line element into the action \eqref{action1}  we get the  following form;
\begin{eqnarray}
S=\frac{1}{2} \int d^4x \,{\cal L} \sqrt{h} &&\Big[{}^{(3)}R + K_{ij} K^{ij} -  K^2 +{\cal G}_{IJ} {\cal L}^{-2} (\dot{\phi}^I-N^i {\phi^I}_{|i}) (\dot{\phi}^J-N^j {\phi^J}_{|j})
\nonumber \\
&&-  {\cal G}_{IJ} {\phi^I}_{|i} \, {\phi^J}^{|i}  -2 V(\phi^K) \Big], \label{action2}
\end{eqnarray}
where $K_{ij} = (N_{i|j}+N_{j|i}-\dot{h}_{ij})/(2{\cal L})$ is extrinsic curvature, $ K=K_{\; i}^i ={\cal L}^{-1}  \left( N^i_{\;\;|i}-\partial_t \ln \sqrt{h} \right)$ is its trace, ${}^{(3)}R $ is three dimensional curvature associated with metric $h_{ij}$,  and spatial indices are raised and lowered by the spatial metric. We also used dot notation for $d/dt$ and vertical bars for three-space-covariant derivatives in terms of a connection compatible with $h_{ij}$. From eq.\eqref{action2} the scalar fields' conjugate momenta are obtained as: 
\begin{equation}
\Pi_I={\cal L}^{-1} {\cal G}_{IJ} (\dot{\phi}^J-N^i {\phi^J}_{|i}).
\end{equation}
Using ADM foliation, Einstein equations are separated into one Hamilton and three momentum constraints plus six dynamical equations for spatial metric.
We may decompose the extrinsic curvature and the spatial metric as follows:  
\begin{eqnarray}
K_{ij}&=& \frac{K}{3} h_{ij}+ a^2(t) \; e^{2 \psi} A_{ij}  ,\\
h_{ij}&=& a^2(t) \; e^{2 \psi} \gamma_{ij},
\end{eqnarray}
where  $A_{ij}$ is traceless, $ \psi $ is curvature perturbation, $a(t)$ is scale factor and
$\gamma_{ij}$ satisfies $\text{det}\gamma_{ij}=1$.  
 We are interested in super-Hubble scales thus it is convenient to expand field equations w.r.t. the spatial derivatives,  called the gradient expansion and associate a parameter $\varepsilon:=k/(aH)$ to each spatial derivative, $ \partial_i={\cal O}(\varepsilon) $. We assume that the universe 
in very large wavelength limit $\varepsilon \rightarrow 0$  is FLRW, i.e.
\begin{equation}
ds^2=-dt^2+ a^2(t) \, \delta_{ij} \, dx^i \, dx^j .
\end{equation}
Therefore we consider $\dot{\gamma}_{ij}={\cal O}(\varepsilon )$ and $N^i={\cal O}(\varepsilon )$. We define effective Hubble rate as the expansion of the unit time-like vector normal to surface of constant $t$, $n^\mu={\cal L}^{-1} (1,-N^i)$. It can be written as:
\begin{equation}
H:=h^{ij} K_{ij}/3=\nabla_\mu n^\mu /3={\cal L}^{-1} (\bar{H}+\dot{\psi} )+{\cal O}(\varepsilon^2) , 
\end{equation}
where $\bar{H}:=\dot{a}/a$.
We use spatially flat gauge $\psi =0 $. Then starting with $N^i, \partial_i, \dot{\gamma}_{ij}$  of the order ${\cal O}(\varepsilon)$ and by using  equation of $\gamma_{ij}$, we obtain the Hamilton constraint as follow;
\begin{equation}
3H^2=\frac{H^2}{2} {\cal G}_{IJ} \phi^I_N \phi^J_N +V + {\cal O}(\varepsilon^2), \label{C8}
\end{equation}
where $\phi^I_N:=d\phi^I/dN$.
The anisotropic stress, i.e. $ T_{ij}-\frac{1}{3} h_{ij} T^k_{\;k} $, vanishes for scalar fields. Consequently, considering $A_{ij}={\cal O}(\varepsilon)$ and using the equation of $A_{ij}$, we can reach to $A_{ij}\varpropto a^{-3} $. Thus we neglect $A_{ij}$ in first order of $\varepsilon$ and consider $A_{ij}={\cal O}(\varepsilon^2)$. On the other hand using dynamical equation of  $\gamma_{ij}$ we can conclude that $\dot{\gamma}_{ij}={\cal O} (\varepsilon^2)$. The equation of scalar fields finally takes the form;
\begin{equation}
H {\cal D}_N (H \phi^I_N) +3H^2 \phi^I_N +{\cal G}^{IJ} V_{,J} + {\cal O}(\varepsilon^2) =0, \label{C9}
\end{equation}
where  the covariant version of  derivatives w.r.t. the number of $e$-folds is defined by ${\cal D}_N=\frac{d\phi^I}{dN} \nabla_I$.
To leading order in $\varepsilon$ the perturbed Hamilton constraint and equation of scalar fields, with a convenient choice of time coordinate, are in the form of background equations. 
Hence, different Hubble patches evolve locally as homogeneous and isotropic FLRW universe; and  independently of one another for local theories. 
We should also consider the momentum constraint;
\begin{equation}
2\, \partial_i H=-H {\cal G}_{IJ}  \phi^I_N \, \partial_i \phi^J + {\cal O} (\varepsilon^3) . \label{C10}
\end{equation}
There is not any equation in background similar to \eqref{C10}.
Thus it is an additional constraint for the separate universe approximation.   By the use of eq.s \eqref{C8} and \eqref{C9} we can also write
\begin{equation}
2\, \partial_i H=-H {\cal G}_{IJ}  \phi^I_N \, \partial_i \phi^J +B_i+  {\cal O} (\varepsilon^3) .
\end{equation}
The existence of $B_i$ makes an error in the separate universe approximation.
However the momentum constraint is satisfied at leading order and $B_i={\cal O}(\varepsilon^3)$ \cite{Sugiyama}.
We can also write 
\begin{equation}
\phi^I_N=2 {\cal G}^{IJ} \frac{H_{,J}}{H}+{\cal O} (\varepsilon^2) , \label{HJ}
\end{equation}
where $H_{,I}:=\partial H/ \partial \phi^I$.
We can define $e$-fold number  locally by
\begin{equation}
{\cal N}(t_2, t_1; x^i):=\frac{1}{3} \int_{t_1}^{t_2} dt^{\prime} \, {\cal L} (t^{\prime},x^i) \, K(t^{\prime},x^i) 
=N(t_2, t_1)+\psi(t_2;x^i)
-\psi(t_1;x^i),
\end{equation} 
where $N(t_2 ,t_1)$ is background number of $e$-folds.
We can choose different sets of slices for times $t_1$ and $t_2$. Considering both the initial and final slices to be flat, we get;
\begin{equation}
{\cal N}(t_2,t_1;x^i)=N(t_2,t_1).
\end{equation}
Also taking the initial slice flat and the final slice uniform energy, we obtain;
\begin{equation}
 {\cal N}(t_2, t_1;x^i)=N(t_2,t_1) +\zeta(t_2;x^i),
\end{equation}
where $\zeta$ is the uniform energy curvature perturbation. Difference between these two above local e-folding numbers is known as $\delta N$;
\begin{equation}
\delta N:={\cal N}(t_2,t_1;x^i)-N(t_2,t_1)=\zeta(t_2;x^i).
\end{equation} 
Scalar-field momenta and Hubble expansion rate at a point are only function of scalar fields at that point. Therefore the field velocities are functions of  scalar fields. It is enough to expand $\delta N $ w.r.t. scalar field perturbations.
 The curvature perturbation on uniform density hypersurfaces after exiting the horizon reads
\begin{equation}
\zeta = \delta N= \sum_I N_{,I}^\ast Q^I +\frac{1}{2} \sum_{IJ}  N_{;IJ}^\ast  Q^I \,  Q^J +\cdots ,
\end{equation}
where $Q^I=\delta\phi^I+ \frac{\dot{\phi}^I}{H} \psi$ are scalar field perturbations in flat gauge.
This formalism enables us to calculate the curvature perturbation  on large scales without solving perturbed field equations.

\section{Flow equations of inflationary parameters}
\label{Flow}

By analogy with single field  inflation, we define
\begin{equation}
\epsilon_I:=\sqrt{2} \frac{H_{,I}}{H}, \label{S1}
\end{equation}
and
\begin{equation}
 \epsilon={\cal G}^{IJ} \epsilon_I \epsilon_J =\frac{2}{H^2} {\cal G}^{IJ} H_{,I} H_{,J}= \frac{1}{2H^2} {\cal G}_{IJ} \dot{\phi}^J \dot{\phi}^J=-\frac{\dot{H}}{H^2} .
\end{equation}
The only condition for having inflation is again $\epsilon < 1$, that corresponds to $ \ddot{a}/a=H^2(1-\epsilon) >0 $.
Note that because of $ \epsilon=\epsilon^I \epsilon_I  $, it is not enough to consider $\epsilon_I < 1$.
The exact form of  potential can be obtained from the Hamilton constraint equation; 
\begin{equation}
 3H^2(\phi^K)=V(\phi^K)+2 \; {\cal G}^{IJ}(\phi^K) \, H_{,I}(\phi^K) H_{,J}(\phi^K).
\end{equation}
Change of $ \epsilon_I$ w.r.t. $e$-folding number can be written as 
\begin{equation}
{\cal D}_N \epsilon_I= {\cal G}^{JK} \epsilon_K \lambda_{IJ} -\epsilon \epsilon_I , \label{flow1}
\end{equation}
here  we have defined the second parameter by
\begin{equation}
\lambda_{IJ}:=2 \frac{H_{;IJ}}{H} ,
\end{equation}
and  $ H_{;IJ}:= \frac{\partial^2 H}{\partial \phi^I \, \partial \phi^J}+\Gamma^K_{IJ} \frac{\partial H}{\partial \phi^K}$ is the covariant derivative w.r.t. scalar fields.
Similarly, it is straightforward to obtain
\begin{equation}
{\cal D}_N \lambda_{IJ}=  {}^{(2)}\lambda_{IJ} -\lambda_{IJ} \epsilon. \label{flow2}
\end{equation}
The parameter ${}^{(2)}\lambda_{IJ} $ and higher order parameters can be written by:
\begin{equation}
{}^{(m)}\lambda_{\nu_0 \, \nu_1}=\left(\frac{2}{H} \right)^m  H_{;\nu_0 \cdots \nu_m} \prod_{i=2}^m H^{, \nu_i} ,
\end{equation}
where we have used $ H^{,I}={\cal G}^{IJ} H_{,J} $. 
Flow equations for these parameters can be obtained as follows:
\begin{equation}
{\cal D}_N {}^{(m)}\lambda_{IJ}=  {}^{(m+1)}\lambda_{IJ} +{}^{(m)}W_{IJ} - m \; { }^{(m)}\lambda_{IJ} \epsilon , \label{flow3}
\end{equation}
where we have used ${\cal G}_{IJ;K}=0$ and defined
\begin{equation}
{}^{(m)} W_{\nu_0 \nu_1}:= \sqrt{2} \left(\frac{2}{H} \right)^m H_{;\nu_0 \cdots \nu_m}  \sum_{\substack{k=2 \\ I=1}}^m \prod_{\substack{j=2 \\ j\neq k}}^{m} \epsilon_I H^{;I  \nu_k} H^{,\nu_j}   .
\end{equation}
The equations \eqref{flow1}, \eqref{flow2} and \eqref{flow3} are a system of first order differential equations.
If for some $m$, all ${ }^{(m)}\lambda_{IJ}$ and ${ }^{(m)}W_{IJ}$  vanish then all the higher order parameters will be zero. This condition corresponds to:
\begin{equation}
H_{;\nu_0 \cdots \nu_m}=0 . \label{T1}
\end{equation}
In single field models the condition \eqref{T1} is reduced to 
\begin{equation}
\frac{d^{m+1}}{d \phi^{m+1}}  H(\phi)=0 .
\end{equation}
Consequently, one can obtain
\begin{equation}
H(\phi)=H_0\left[ 1+ \sum_{n=1}^{m} G_n \; (\phi)^n \right] ,
\end{equation}
where $G_n$ are integration constants.
With choosing initial values one can obtain the full inflationary dynamics. In similar manner in the flat field space one can conclude that $ H=\sum_{I,J,\cdots} P_{IJ\cdots} \left({\phi^1}\right)^I \, \left( {\phi^2}\right)^J \, \cdots $ where $I+J+\cdots=0 $ up to  $ I+J+\cdots=m $. In general form of curved field space the condition is more complicated.
Thus vanishing higher order covariant derivatives of Hubble expansion rate leads to a more complicated form.
We can recap equations of scalar fields by the use of the parameters as follow
\begin{equation}
\epsilon^K \lambda_{JK} - 3\epsilon_J+ {\cal G}_{JI}  \Gamma^I_{MK} \epsilon^M \epsilon^K + \frac{V_{,J}}{\sqrt{2} H^2}=0 . \label{C7}
\end{equation}
In slow-roll regime all dynamical  properties of the universe change a little over a single $e$-folding of expansion, thus first and third terms can be neglected.
Considering $\epsilon_I$, $\lambda_{IJ} \ll1$,  equation \eqref{C7} becomes
\begin{equation}
\epsilon_I=\frac{1}{\sqrt{2}} \frac{V_{,I}}{V} ,
\end{equation}
where we have used $ 3H^2=V $.
It is common to define slow-roll parameters  by the use of potential and its derivatives,
\begin{equation}
\epsilon_{IJ}:=\frac{1}{2} \frac{V_{,I} V_{,J}}{V^2},
\qquad
\epsilon=\mathrm{tr} \, \epsilon_{IJ},
\qquad
\eta_{IJ}:= \frac{V_{;IJ}}{V}.
\end{equation}
Consequently, the slow-roll parameter have the relation with previous parameter as follow
\begin{equation}
\epsilon_I=\frac{1}{\sqrt{2}}\frac{V_{,I}}{V} ,
\qquad
\epsilon_{IJ}=\epsilon_I \epsilon_J ,
\qquad
\eta_{IJ}=\lambda_{IJ}+\epsilon_I \epsilon_J .
\end{equation}
Notice that the condition of the end of inflation $\epsilon(H)=1$ is exact whereas $\epsilon(V)=1$ is an approximation.

\section{Cosmological perturbations}
\label{CP}

In this section we try to obtain the quantities that we need in $\delta N$ formalism.
 Assuming linear scalar perturbations about the homogeneous metric as
\begin{equation}
ds^2= -(1+2A) dt^2 +2a (\partial_i B ) dx^i \, dt +a^2 [(1-2\psi) \delta_{ij} + 2 \partial_i \partial_j E] \, dx^i \, dx^j,
\end{equation}
the scalar fields' action, i.e. the last two terms in action \eqref{action1}, in second order of perturbations take the form of 
\begin{equation}
S^{(2)}=\int d^4x \, a^3 \left( {\cal D}_t Q_I\, {\cal D}_t Q^I-h^{ij}\, \partial_i Q_I \, \partial_j Q^I - {\cal M}_{IJ} Q^I Q^J  \right),
\end{equation}
where the effective mass matrix has the following form:
\begin{equation}
{\cal M}_{IJ}=V_{;IJ} - R_{KIJM} \dot{\phi}^K \dot{\phi}^M - \frac{1}{a^3} {\cal D}_t \left( \frac{a^3}{H} \dot{\phi}_I \dot{\phi}_J \right).
\end{equation}
By the use of field equations
Sasaki and Stewart calculated the power spectrum after horizon-exit,
\begin{equation}
\langle Q^I({\bf k}_1) Q^J({\bf k}_2) \rangle=(2 \pi)^3 \, \delta^{(3)}({\bf k}_1 + {\bf k}_2) \frac{H^2}{2k^3} {\cal G}^{IJ}.
\end{equation}
where $Q^I_{\bf k}$ is the Fourier transformed of $Q^I$, $\delta^{(3)}$ is three dimensional Dirac delta function and ${\bf k}$ are comoving wave-numbers.
Using $\delta N$-formalism, power spectrum of the curvature perturbation is then defined by 
\begin{equation}
\langle \zeta_{\mathbf{k}_1}  \zeta_{\mathbf{k}_2} \rangle := (2\pi)^3 \delta({\bf k}_1+{\bf k}_2) \, P_\zeta (k_1) .
\end{equation}
The dimensionless power spectrum  $ {\cal P}_\zeta(k)=\frac{k^3}{2\pi^2} P_\zeta(k) $ becomes
\begin{equation}
{\cal P}_\zeta(k)= \frac{H^2}{4\pi^2} {\cal G}^{IJ} N_{,I} N_{,J}  ,
\end{equation}
and spectral index, that describes  deviation from scale invariance, takes the following form:
\begin{equation}
n_s-1=\frac{d\ln {\cal P}_\zeta}{d \ln k}=2\frac{\dot{H}}{H^2}- 2 \frac{N_{,I} \left( \phi^I_N \phi^J_N +\frac{1}{3} R^{I\quad J}_{\;KL}  \phi^K_N \phi^L_N - \frac{V^{;IJ}}{V} \right) N_{,J}}{{\cal G}^{OP} N_{,O} N_{,P}}.
\end{equation}
Some parts of curvature perturbation can be generated at the hypersurface of the end of inflation $ \zeta_c $ in addition to the curvature perturbation generated during inflation $\zeta_{inf}$. 
In multiple field inflation the hypersurface of the end of inflation can be different from uniform energy hypersurface, e.g. in two field hybrid inflation the hypersurface of the end of inflation defines an ellipse. Thus the curvature perturbations has two parts $ \zeta=\zeta_{inf}+\zeta_c$.
 However we can consider situation that $\zeta \simeq \zeta_{inf}$.
We consider 
the time $t_\ast$  to be flat initial hypersurface, whereas $t_c$  the uniform energy density hypersurface and
the final time $t_e$ to be a time deep inside radiation dominated era when reheating is completed. By use of $ \rho=\rho_0 a^{-4} $ we can obtain
\begin{equation}
N_c=N(t_c)-N(t_e)=\frac{1}{4} \ln \left(\frac{\rho_e}{\rho_c} \right) = \frac{1}{2} \ln \left(\frac{H^e}{H^c} \right),\label{end}
\end{equation}
however if the end of inflation coincide with uniform density hypersurface then $ \delta N_c=0 $. By the use of eq.\eqref{end}, we can write  the perturbation in the number of $e$-folds due to the end of inflation as
\begin{equation}
\delta N_c=\frac{1}{2 H^e} \sum_{IJ} H_{,J}^e \left( \frac{\partial \phi_e^J}{\partial \phi_\ast^I}\right) \, Q^I_\ast ,
\end{equation}
or
\begin{equation}
\delta N_c=\frac{1}{2\sqrt{2}} \sum_{IJ} \epsilon^e_J \left( \frac{\partial \phi_e^J}{\partial \phi_\ast^I}\right) \, Q^I_\ast , \label{C6}
\end{equation}
where we have used the fact that curvature perturbation in radiation era is constant. 
Now we try to obtain $\delta N_{inf}$.
In multifield inflation there are many possible inflationary trajectories. The initial conditions are important in choosing the true inflationary trajectory and the position of inflaton on it.
It is common in such cases that find a \emph{constant of motion}.
 We define a vector $B_I$ in field space and  quantity $C_\gamma:=\int_\gamma B_I d\phi^I$, which should be independent of the particular path $\gamma$ and constant along trajectory, thus
\begin{equation}
C_\gamma =\int_\gamma B_I d\phi^I= \int_\gamma \Phi_{,I} d\phi^I.
\end{equation}
Vanishing of  derivative of $C_\gamma$ leads to 
\begin{equation}
  \Phi_{,I} H^{,I}=0 .
\end{equation}
We define arbitrary functions of $\alpha_I (\phi^J)$ as follow
\begin{equation}
\Phi_{,1}=\frac{\alpha_1(\phi^J)}{H^{,1}}, \quad 
\Phi_{,I}=\frac{\alpha_I(\phi^J) - \alpha_{I-1}(\phi^J)}{H^{,I}}, \quad 
\Phi_{,M}=\frac{ - \alpha_{M-1}(\phi^J)}{H^{,M-1}},
\end{equation}
where $I$ goes from two up to $M-1$.
Then $C_\gamma$ becomes
\begin{equation}
C_\gamma = \int \left[\alpha_1\left( \frac{d\phi^1}{H^{,1}}-\frac{d\phi^2}{H^{,2}} \right)
 +\alpha_2 \left( \frac{d\phi^2}{H^{,2}}-\frac{d\phi^3}{H^{,3}} \right)
  +\cdots +\alpha_{M-1} \left( \frac{d\phi^{M-1}}{H^{,M-1}}-\frac{d\phi^M}{H^{,M}} \right) \right] .
\end{equation}
Consequently we define 
\begin{equation}
C_I:=\int \alpha_I(\phi^J) \, \left( \frac{d\phi^I}{H^{,I}}-\frac{d\phi^{I+1}}{H^{,I+1}} \right)
\quad
\text{ (no sum on I)}
, \qquad C_\gamma= \sum_{l=1}^{M-1} C_l. \label{C1}
\end{equation}
Using above quantities we can obtain different constants for different set of $\alpha_I$. 
Choosing the same $\alpha_I$ for all $I$ leads to  $C_\gamma=\sum_{L=1}^{M-1} C_l=\int \alpha \left( \frac{d\phi^1}{H^{,1}} - \frac{d\phi^M}{H^{,M}} \right)$. If we use $\alpha_I=\alpha$ for   $I \leq K$ and vanishing $\alpha_I$  for the other cases we get  $\tilde{C}_K:=\sum_{L=1}^{K-1} C_L=\int \alpha \left( \frac{d\phi^1}{H^{,1}} - \frac{d\phi^M}{H^{,M}} \right) $.
By use of the same non-vanishing $ \alpha_I $ for  $ I \le K \le J $ we reach at $ C_{I,J}  := \int \alpha \left( \frac{ d \phi^I}{H^{,I}} - \frac{d\phi^J}{H^{,J}} \right) $.
We can also define other constants, but we do not use them in this work. 
If we wish derivatives of $C_I$ depend only on the choice of trajectory, then $\Phi_{,I}$ needs to be only a function of $\phi^I$ i.e. $(\Phi_{,I})_{,J}=0 $ for $I\neq J$.
However in definition of $C_I$, i.e. eq.\eqref{C1}, we have the non-vanishing quantities by $\Phi_{,I}=\frac{\alpha_I}{H^{,I}}$ and $\Phi_{,I+1}=-\frac{\alpha_I}{H^{,I+1}}$. Thus we  obtain
\begin{equation}
\left(\frac{\alpha_I}{H^{,I}} \right)_{,J}=0 \qquad {\rm for} \qquad I \neq J,
\end{equation}
and 
\begin{equation}
\left(\frac{\alpha_I}{H^{,I+1}} \right)_{,J}=0 \qquad {\rm for} \qquad I+1 \neq J.
\end{equation}
To employ $\delta N$-formalism we have to evaluate derivative of the number of $e$-folds w.r.t. scalar fields on initial hypersurface $t_\ast$.
For this reason we write the number of $e$-folds  in general form  by
\begin{equation}
N=\int_\ast^e H  \, dt=-\frac{1}{2} \sum_I \int_\ast^e A_I \, d\phi^I ,
\end{equation}
where
\begin{equation}
H=-\frac{1}{2} \sum_I A_I \frac{d\phi^I}{dt}=\sum_I A_I H^{,I} .
\end{equation}
 We need derivative of the $e$-folding number w.r.t. scalar fields,
\begin{equation}
\frac{\partial N}{\partial \phi^M_\ast}=\frac{1}{2} \sum_L  \left[ A^\ast_L \delta^L_M - A^e_L \frac{\partial \phi_e^L}{\partial \phi_\ast^M} - \sum_{K \neq L} \int_\ast^e (A_L)_{,K} \frac{\partial \phi^K}{\partial \phi^M_\ast} \, d\phi^L \right]. \label{C12}
\end{equation}
Assuming ans\"{a}tz $(A_L)_{,K}=0$ for $K\neq L$,  eq.\eqref{C12} reduces to
\begin{equation}
\frac{\partial N}{\partial \phi^M_\ast} = \frac{1}{2} \sum_L  \left[ A^\ast_L \delta^L_M - A^e_L \frac{\partial \phi_e^L}{\partial \phi_\ast^M} \right] . \label{N1}
\end{equation}
Our work then becomes calculating  quantity $ \frac{\partial \phi^M_e}{\partial \phi^K_\ast} $.
To use eq.\eqref{N1} what we need is
\begin{equation}
\frac{\partial \phi^M_c}{\partial \phi^K_\ast}=\sum_L \frac{\partial \phi^M_c}{\partial C_L} \frac{\partial C_L}{\partial \phi^K_\ast}.
\end{equation}
Using eq.\eqref{C1} we can obtain derivative of $C_I$ w.r.t. scalar fields on initial hypersurface
\begin{equation}
\frac{\partial C_I}{\partial \phi^L_\ast}=\left(\frac{\alpha_I}{H^{,I}}\right)^\ast \delta^I_L - \left(\frac{\alpha_I}{H^{,I+1}}\right)^\ast \delta^{I+1}_L =\left(\frac{\alpha}{H^{,L}}\right)^\ast ( \delta^I_L- \delta^I_{L-1}), \label{C4}
\end{equation}
where we have used  $\alpha_I=\alpha$.  We can write
\begin{equation}
\frac{\partial C_L}{\partial C_K}=\delta^L_K = \frac{\alpha}{H^{,L}} \frac{\partial \phi^L}{\partial C_K} - \frac{\alpha}{H^{,L+1}} \frac{\partial \phi^{L+1}}{\partial C_K} . \label{C2}
\end{equation}
and also
\begin{equation}
\frac{\partial \tilde{C}_K}{\partial C_L} = \left( \frac{\alpha}{H^{,1}} \right)^e \frac{\partial \phi^1_e}{\partial C_L} - \left( \frac{\alpha}{H^{,K}} \right)^e \frac{\partial \phi^K_e}{\partial C_L}.
\end{equation}
The condition for the hypersurface of the end of inflation can be written by
\begin{equation}
E(\phi^I_e)=const
\end{equation}
and consequently we get
\begin{equation}
\sum_L E^e_{,L} \frac{\partial \phi^L_e}{\partial C_K}=0 .
\end{equation}
If the end of inflation coincide with a uniform  hypersurface we can put $E=H$. By rewriting eq.\eqref{C2} as 
\begin{equation}
\frac{1}{H^{,1}} \frac{\partial \phi^1_e}{\partial C_K} =\frac{\sum_{L=K+1}^{N} E^e_{,L} H^{e ,L} }{\alpha^e \sum_{L=1}^{N} E^e_{,L} H^{e ,L}},
\end{equation}
we have
\begin{equation}
\frac{\partial \phi^K_e}{\partial C_L}= \left( \frac{H^{,K}}{\alpha} \right)^e \left[\frac{\sum_{M=L+1}^{N} E^e_{,M} H^{e ,M} }{\sum_{M=1}^{N} E^e_{,M} H^{e ,M}}- \Theta_{LK} \right], \label{C3}
\end{equation}
where we have  defined $ \Theta_{LK}=\frac{\partial \tilde{C}_K}{\partial C_L} $ that is 1 for $ K \leq L-1 $ and vanishes for $ K > L-1 $.
Then eqs.\eqref{C4} and \eqref{C3} lead to
\begin{equation}
\frac{\partial \phi^L_e}{\partial \phi^M_\ast}= \left(\frac{H^{,L}}{\alpha}\right)^e \left(\frac{\alpha}{H^{,M}}\right)^\ast \left[ \frac{E^e_{,M } H^{e ,M} }{\sum_{S=1}^{N} E^e_{,S} H^{c,S}} -\Theta_{M-1,L} + \Theta_{M,L} \right]. \label{C5}
\end{equation}
Having eq.\eqref{C5} we can obtain $ \left( \frac{\partial N}{\partial \phi^I_\ast} \right)_{inf} $ as follows:
\begin{equation}
\left( \frac{\partial N}{\partial \phi^I_\ast} \right)_{inf}=\frac{1}{2} A^\ast_I +\frac{1}{2} \left( \frac{\epsilon^I H}{\alpha} \right)^e \left( \frac{\alpha}{\epsilon^I H} \right)^\ast \left[ A_I - \sqrt{2} \frac{\bar{\epsilon}_I \epsilon^I}{\bar{\epsilon}}\right]^e .
\end{equation}
Plugging eq.\eqref{C5} into eq.\eqref{C6} we can also obtain
\begin{equation}
\left( \frac{\partial N}{\partial \phi^I_\ast} \right)_{c}=\frac{1}{2\sqrt{2}} \left( \frac{\epsilon^I H}{\alpha} \right)^e \left( \frac{\alpha}{\epsilon^I H} \right)^\ast \left[  \frac{\bar{\epsilon}_I \epsilon^I}{\bar{\epsilon}} - \epsilon_I \right]^e ,
\end{equation}
where we have defined $\bar{\epsilon}_I :=\sqrt{2} E_{,I}/ H$ and $\bar{\epsilon} := \sum_I \bar{\epsilon}_I \epsilon^I $.
Now we want to show how the ans\"{a}tz restricts the form of Hubble expansion rate. We consider a simple two field curved model with line element $ d\phi^2+ e^{-2b(\phi)} \, d\psi^2 $, where $b(\phi)$ is a function of $\phi$. This model can be motivated from transforming   $f(R)$ minimally couple with a canonical scalar field into Einstein's frame.
We can also define $ F_I:=A_I H^{,I} $ without sum on $I$ indices, thus $A_I=\frac{F_I(\phi^J)}{H^{,I}} $, 
where $F_I$ are arbitrary functions of scalar fields, we will restrict them by the ans\"{a}tz. 
 The anst\"{a}tz $ \left( \frac{F_I}{H^{,I}}\right)_{,J}=0 $ with $I \neq J$ in the two fields model is reduced to
\begin{equation}
\left( \frac{F_\phi}{H_{,\phi}}\right)_{,\psi}=0=\left( \frac{F_\psi}{H_{,\psi}} e^{-2b}\right)_{,\phi}, \qquad  {\rm with} \qquad F_\phi +F_\psi =H. \label{A1}
\end{equation}
Thus we can write $F_\phi=A(\phi) \, H_{,\phi} $ and $F_\psi=B(\psi) \, e^{-2b} H_{,\psi} $, where $A$ and $B$ are functions of $\phi$ and $\psi$, respectively. If we consider Hubble parameter  of the form of $H=H(\omega)$ where $\omega=\left[u(\phi)\right]^\gamma \, \left[v(\psi)\right]^\beta$ is product separable, then we have $H=H_\omega \omega (\gamma+\beta e^{-2\omega})$ where we have set
\begin{equation}
A=\frac{u}{u_\phi}, \qquad B=\frac{v}{v_\psi} e^{-2v^\beta} \qquad {\rm and} \qquad b=u^\gamma.
\end{equation}
On the other hand, the second assumption $(\alpha/H^{,I})_{,J} =0$ with $I \neq J$ gets the following form:
\begin{equation}
\left( \frac{\alpha}{H_{,\phi}} \right)_{,\psi}=0=\left( \frac{\alpha}{H_{,\psi}} e^{-2 b(\phi)} \right)_{,\phi}. \label{A2}
\end{equation}
By solving  eq.\eqref{A2} we obtain $\alpha=c H_\omega \omega e^{2 b(\phi)}$, where c is an integration constant.
If we assume $H=H(\omega)$ with $\omega= \gamma \, u(\phi) + \beta \, v(\psi)$ then eq.\eqref{A1} leads to $H=H_\omega (\gamma + \beta e^{-2 \omega} )$ where we set
\begin{equation}
A=\frac{1}{u_\phi}, \qquad B=\frac{e^{-2\beta v}}{v_\psi} \qquad {\rm and} \qquad b= \gamma u.
\end{equation}
And eq.\eqref{A2} leads to $\alpha= c H_\omega e^{2 b(\phi)}$.

\section{Kinematic basis}
\label{The adiabatic direction}

In multi-field inflation the comoving curvature perturbation is defined as follow
\begin{equation}
{\cal R}=\psi-H \sum_I \frac{\dot{\phi}_I}{\sum_J \dot{\phi}^J \dot{\phi}_J} \, \delta \phi^I=-H \sum_I \frac{\dot{\phi}_I}{\sum_J \dot{\phi}^J \dot{\phi}_J} Q^I .
\end{equation}
In Hamilton-Jacobi (HJ) approach this relation becomes $ {\cal R}= \overrightarrow{\nabla}H.\overrightarrow{Q}/\epsilon H^2.
 $
Thus the comoving curvature perturbation is  the component of scalar fields perturbation in direction of gradient of the Hubble expansion rate.\footnote{In slow-roll  approximation this relation can be written by potential, i.e. ${\cal R} \simeq \epsilon^{-1} \overrightarrow{\nabla}V.\overrightarrow{Q}/V $.} Thus it is more convenient to rotate field space and write quantities in the basis of  instantaneous adiabatic $\hat{\sigma}^I= \frac{\dot{\phi}^I}{\dot{\sigma}} $  and entropic directions characterised by $\hat{s}_I^{\; J}=\delta_I^J-\hat{\sigma}^J \hat{\sigma}_I $.
The quantities $\hat{\sigma}^I$ and $\hat{s}^{IJ}$ satisfy following relations
\begin{eqnarray}
\hat{\sigma}^I \hat{\sigma}_I &=& 1, \nonumber \\
\hat{s}^{IJ} \hat{s}_{JK} &=& \hat{s}^{I \,K}, \nonumber \\
\hat{\sigma}_I \hat{s}^{IJ} &=& 0 .
\end{eqnarray}
 The rate of change of $ \hat{\sigma}^I $ is shown by turn-rate \cite{Peterson},
\begin{equation}
\omega^I:={\cal D}_t \hat{\sigma}^I= -\frac{1}{\dot{\sigma}} V_{,J} \hat{s}^{IJ},
\end{equation}
where we have $\omega^I \hat{\sigma}_I =0$.
We can decompose any vector by use of $\hat{\sigma}^I$ and $\hat{s}^{IJ}$
\begin{equation}
A^I=\hat{\sigma}^I \hat{\sigma}_J A^J +\hat{s}^I_{\, J} A^J.
\end{equation} 
The background equations take the following form in kinematic basis:
\begin{eqnarray}
3H^2 &=&\frac{\dot{\sigma}^2}{2} +V , \nonumber \\
-2 \dot{H} &=& \dot{\sigma}^2,
\end{eqnarray}
and
\begin{equation}
\ddot{\sigma}+ 3 H \dot{\sigma} + V_\sigma = 0, \qquad V_\sigma= V_I \hat{\sigma}^I.
\end{equation}
The background equations are similar to the single field case. However, potential depends on all scalar fields.
One can show in H.J. approach
\begin{equation}
H_{,I}= \left(  \hat{\sigma}^J \hat{\sigma}_I + \hat{s}_I^{\; J} \right) H_{,J}= \hat{\sigma}_I H_{,\sigma}
\quad 
\text{and}
\quad
H_{,\sigma} = -\frac{1}{2} \dot{\sigma} ,
\end{equation}
where $\sigma$ is the integrated  path length along the trajectory and we have used $ H_{,I} \varpropto \hat{\sigma}_I $.
Note that in background level $s=\dot{s}=\ddot{s}=0$.\footnote{By Stewart-Walker lemma the entropic perturbation $\delta s$ is automatically gauge invariant.}
The exact equation $H_{,s}=-\dot{s}/2=0$ will simplify the flow equations.\footnote{We can also write $\dot{s}/H \simeq -M_{pl}^2 V_{,s}/V$ and  $V_{,s} \simeq 0$ on slow roll trajectory.} The parameters of inflation in new basis can be obtained as
$\epsilon_\sigma=\sqrt{2} H_{,\sigma} / H$ and $\epsilon_s=\sqrt{2} H_{,s} / H=0$. Thus $\epsilon={\epsilon_\sigma}^2$ and ${\cal D}H/dN = {\epsilon_\sigma}^2 H$.
Change of $\epsilon_\sigma$ w.r.t. number of $e$-folds becomes:
\begin{equation}
\frac{{\cal D} \epsilon_\sigma}{d N}=\lambda_{\sigma \sigma} \epsilon_\sigma -\epsilon \epsilon_\sigma ,
\end{equation}
where we have defined $\lambda_{\sigma \sigma}=2H_{, \sigma \sigma}/ H$. Similarly we can obtain
\begin{equation}
\frac{{\cal D}\lambda_{\sigma \sigma}}{dN}= { }^{(2)}\lambda_{\sigma \sigma} - \epsilon \lambda_{\sigma \sigma},
\qquad \text{ with} \quad { }^{(2)}\lambda_{\sigma \sigma}:=4\frac{H_{,\sigma} H_{,\sigma \sigma \sigma}}{H^2}.
\end{equation}
Higher order parameters are defined as:
\begin{equation}
{ }^{(m)}\lambda_{\sigma \sigma} = \left( \frac{2}{H} \right)^m (H_{,\sigma})^{m-1} \, \partial^{m+1}_\sigma H .
\end{equation}
we can finally write 
\begin{equation}
\frac{{\cal D} { }^{\; (m)}\lambda_{\sigma \sigma}}{dN} =  { }^{(m+1)}\lambda_{\sigma \sigma}- m \epsilon { }^{(m)}\lambda_{\sigma \sigma} + {}^{(m)}W_{\sigma \sigma}.
\end{equation}
 Thus the hierarchy equations of inflation parameters can be written as if there exists one field.
We can write the number of $e$-folds as follows:
\begin{equation}
N=\int_{\sigma^\ast}^{\sigma^e(t^\ast)} \frac{H}{\dot{\sigma}} \, d\sigma \label{H1} ,
\end{equation}
where the upper limit depends on the time of horizon crossing. We have also used $ N= \int H dt =  \int \sum_I A_I\, ds^I + \int A_\sigma \, d \sigma $ and $\dot{s}^I=0$.
 In first order of perturbation, we can use \eqref{H1} to write \cite{Tye}
\begin{equation}
\delta N= -\frac{H}{\dot{\sigma}} \Big\vert_{t^\ast} Q^\sigma + \frac{H}{\dot{\sigma}} \Big\vert_{t^e} (\partial_I \sigma^e) Q^I - \int_{t^\ast}^{t^e} \frac{2 H}{\dot{\sigma}} \omega^I  Q_I \, dt \label{Ab1} .
 \end{equation}
 As pointed out in ref.\cite{Tye}
 first term corresponds to adiabatic perturbation in single field case.
 The second term arises when
the hyper-surface of the end of inflation is not orthogonal to the background inflaton path.
The last term only contains isocurvature perturbations,
as it represents perturbations orthogonal to the integral path.
 In super-horizon scales ($  k \ll a H $) we have \cite{Multifield3}
 \begin{equation}
 \dot{\zeta} = -\frac{2 H}{\dot{\sigma}} \omega^I  Q_I.
 \end{equation}
There are $N-1$ entropy modes. However only a component of entropy perturbations is source of $\dot{\zeta}$ outside of horizon.
Expanding $\delta N$ we can write
 \begin{equation}
\delta N = N_\sigma Q^\sigma +  N_I \delta s^I, \label{Ab2}
 \end{equation}
 where we have used $ Q_\sigma=\hat{\sigma}^I Q_I $ and $ \delta s^I=\hat{s}^{IJ} Q_J $. 
 Comparing eq.\eqref{Ab1} with eq.\eqref{Ab2}  and neglecting effects of the end of  inflation we get
 \begin{eqnarray}
 N_\sigma &=& -\frac{H}{\dot{\sigma}} \vert_{t^\ast} , \nonumber \\ 
 N_J\hat{s}^J_I &=&  -\int_{t^\ast}^{t^e} \frac{2H}{\dot{\sigma }} \omega_I \, dt .
 \end{eqnarray}
 We can see that $N_J\hat{s}^J_I$ depends on whole inflaton's trajectory. However, this term vanishes for a straight trajectory, and only the existence of turn in the trajectory is its source. To calculate such integration we need to know full inflaton's trajectory.
 In the case that the trajectory is a straight line the adiabatic and the entropic modes remain decoupled, and the density perturbation only comes from the adiabatic perturbation.
 As we have mentioned before, to use $\delta N$-formalism, the final hypersurface should be uniform. Considering the effect of the end of inflation we should use the same idea of previous section and put the final hypersurface deep inside radiation dominated era. Finally $\delta  N$-formalism can be obtained by $\zeta=\delta N + \delta N_c$. 
 
 \section{Non-minimal coupling}
\label{Non-minimal coupling}

Up to now we have worked in Einstein's frame in which the scalar fields coupled minimally to gravity. The nonminimal coupling, however, can exist for some reasons. 
The more general coupling of scalar fields and gravity in D-dimension can be written as follow
\begin{equation}
S_{Jordan}=\int d^Dx \, \sqrt{-g} \left[ \frac{1}{2} \, f(\phi^I,R) - \frac{1}{2} {\cal G}_{IJ} \partial_\mu \phi^I \, \partial^\mu \phi^J - V(\phi^I) \right]. \label{Jordan}
\end{equation}
Using a convenient conformal transformation and adding a new scalar field that appears due to non-linearity of $f(\phi^I,R)$ w.r.t. Ricci scalar, one can write action \eqref{Jordan} in Einstein frame as follow
\begin{equation}
S_{Einstein}=\int d^Dx \, \sqrt{-\hat{g}} \left[ \frac{M_{(D)}^{D-2}}{2} \, R - \frac{1}{2} \hat{\cal G}_{IJ} \hat{\partial}_\mu \phi^I \, \hat{\partial}^\mu \phi^J - \hat{U} (\phi^I) \right], \label{Einstein}
\end{equation}
where we use $\phi^I= \phi^1, \phi^2, \dots \phi^M, \chi$ for new set of scalar fields. New metric of field space and potential get following form, respectively, 
\begin{eqnarray}
\hat{\mathcal{G}}_{IJ} &=& \frac{M^{D-2}_{(D)}}{f^2_{,\chi}} \Bigg[ f_{,\chi} \mathcal{G}_{IJ} + \left( \frac{D-1}{D-2} \right)  f_{,\chi I} f_{,\chi J}  \Bigg], \quad
\text{ with} \quad
\mathcal{ G}_{I\chi}=\mathcal{ G}_{\chi I}=0=\mathcal{G}_{\chi \chi},
\nonumber \\ 
\hat{U}(\phi^I)& =& \left( \frac{M_{(D)}^{D-2}}{f_{,\chi}} \right)^{\frac{D}{D-2}} \left[ V(\phi^I)-f(\phi^I,\chi)+f_{,\chi}(\phi^I,\chi) \, \chi \right].
\end{eqnarray}
And $M_{(D)}^{D-2}$ is reduced Planck mass in D-dimension.
In order to avoid the ghost instability, the eigenvalues of $\hat{{\cal G}}_{IJ}$ should be positive.
The non-minimal coupling induces a curved field space in Einstein's frame.
We just write the relation \eqref{HJ} for Einstein's action \eqref{Einstein}. Moreover transforming eq.\eqref{HJ} into Jordan's frame leads to
\begin{equation}
H_{,I}=-\frac{1}{4f_{,\chi}^2} \left( {\cal G}_{IJ} + f_{,\chi I} f_{, \chi J} \right) \dot{\phi}^J + \frac{1}{2f_{,\chi}} \left[ f_{,\chi I} \left( H-\frac{d{f_{,\chi}}/dt}{2f_{, \chi}} \right) -(d{f_{, \chi}/dt)_{, I}} \right],
\end{equation}
where it reduces to common relation of first section for constant values of  $f_\chi $.
Using such relation in defining inflationary parameters leads to a set of more complicated flow equations that we do not study in this work.
Conformal transformation just relabels the metric.
Mathematically at classical level  cosmological observables are not altered by this transformations\cite{ob,ob2,ob3}. Consequently we are free to perform the transformation and work in either frames. 
In general, adiabaticity and evolution of the curvature perturbation is frame-dependent.
Note that the curvature perturbation is not an observable and the physical observables must be carefully defined.

\section{Conclusion}

In this work we have considered multi-scalar fields to derive the evolution of inflationary epoch. We have written the flow equations of inflationary parameters in curved field space. Due to covariant derivatives the condition for truncating the equations is more complicated than flat field space. Consequently, one cannot obtain the Hubble parameter in general form, analytically. We then obtained the derivatives of the number of $e$-folds w.r.t. scalar fields used in $\delta N$-formalism during and at the end of inflation. We finally wrote fields in adiabatic and entropic basis. The behavior of flow equation in new basis is just similar to the single field model, but however depends on all degrees of freedom.

\end{document}